\documentclass{PoS}

\def\tauKpi{$\tau\to K \pi \nu_\tau$}

\def\beq {\begin{equation}}
\def\eeq {\end{equation}}
\def\bea {\begin{eqnarray}}
\def\eea {\end{eqnarray}}

\def\rd{\mbox{d}}
\def\IK{I_{K_{e3}^0}}

\def\IM {\mbox{Im}}
\def\RE {\mbox{Re}}

\newcommand{\sand}[3]{\langle#1|#2|#3 \rangle}

\begin{boldmath}
\title{Dispersive representation of the $K\pi$ vector form factor  and fits to \tauKpi~and $K_{e3}$ data }
\end{boldmath}

\ShortTitle{K$\pi$  vector form factor:  dispersive representation and fits to \tauKpi~and $K_{e3}$ data}

\author{\speaker{Diogo R. Boito}\\
Grup de F\'{\i}sica Te\`orica and IFAE,\\ Universitat
 Aut\`onoma de Barcelona, E-08193 Bellaterra (Barcelona), Spain.\\
              E-mail: \email{boito@ifae.es}}
\author{Rafel Escribano\\
        Grup de F\'{\i}sica Te\`orica and IFAE,\\ Universitat
 Aut\`onoma de Barcelona, E-08193 Bellaterra (Barcelona), Spain.\\
        E-mail: \email{escribano@ifae.es}}

\author{Matthias Jamin\\
        Instituci\'o Catalana de Recerca i Estudis Avan\c{c}ats (ICREA), Grup de F\'{\i}sica Te\`orica and IFAE,\\ Universitat
 Aut\`onoma de Barcelona, E-08193 Bellaterra (Barcelona), Spain.\\
        E-mail: \email{jamin@ifae.es}}

\abstract{Recently, we introduced several  dispersive
  representations for the vector $K\pi$ form factor  and fitted them to
  the Belle spectrum of \tauKpi. Here, we briefly present the model
  and discuss the results for the slope and curvature of $F_+(s)$
  arising from the best fit. Furthermore, we compare the pole position of
  the charged $K^*(892)$ computed from our model with other results in
  the literature. Finally, we discuss the prospects of a simultaneous
  fit to  \tauKpi~and  $K_{e3}$ spectra. }

\FullConference{International Workshop on Effective Field Theories: from the pion to the upsilon\\
                2-6 February 2009\\
                Valencia, Spain}

\begin{document}


\section{Introduction}

Decays of the $\tau$-lepton into hadrons are an important source of
information about a wealth of fundamental parameters in the standard
model. An important example is the QCD coupling $\alpha_s$ that can be
extracted from inclusive $\tau$ decays~\cite{alphas}. After the
separation of Cabibbo-allowed and Cabibbo-suppressed decay modes into
strange particles, the mass of the strange quark and the quark-mixing
matrix element $|V_{us}|$ could  also be determined~\cite{msVus}. More
recently, the $B$-factories have gathered high-statistics data for
exclusive channels. In this work we deal with  $\tau\to K_S\pi \nu_\tau$ decays for
which a spectrum  became available from Belle~\cite{Belle}.  In
these decays, the $K\pi$ form factors can be studied.  Furthermore, the
isolated hadronic pair in the final state constitutes a clean
environment to the study of $K\pi$ interactions.  Therefore,
information about $K\pi$ resonances can also be obtained.

The $K\pi$ form factors are key ingredients in the benchmark
extraction of $|V_{us}|$ from $K_{l3}$ decays~\cite{LR}.  They are defined as follows \cite{Flavianet}
\beq
\sand{\pi^-(p)}{\bar s\, \gamma^{\,\mu} \,u}{K^0(k)}=       \left[  (k+p)^\mu  - \frac{m_K^2 -m_\pi^2}{q^2}(k-p)^\mu\right]F_+(q^2)   + \frac{m_K^2 -m_\pi^2}{q^2}(k-p)^\mu F_0(q^2)  , \label{FF}
\eeq
where $F_+(q^2)$ and $F_{0}(q^2)$ are  the vector and scalar form
factors respectively and  $ q^2 = (k-p)^2$. It follows from the definition that at $q^2=0$ we have $F_+(0)=F_0(0)$. It is then convenient to work with normalised
form factors $\tilde F_{+,0}(q^2)$ such that 
\beq
F_{+,0}(q^2)=F_{+,0}(0)\tilde F_{+,0}(q^2).
\eeq
On the one hand, a reliable value for the normalisation at zero is
crucial in order to disentangle  the product $|V_{us}|F_{+,0}(0)$ that can be extracted from $K_{l3}$ decays. In this respect, chiral perturbation theory  and
lattice QCD are the most trustworthy methods to obtain $F_{+,0}(0)$. On
the other, the energy dependence  of the form factors, encoded in $\tilde F_{+,0}(q^2)$, is needed when
performing the phase space integrals for $K_{l3}$ decays. Here, 
we tackle  the latter aspect of the problem.

In the context of $K_{l3}$ decays, where $m_l^2<t\equiv q^2<(m_K-m_\pi)^2$
  it is customary to   Taylor expand the form factors
\beq
\tilde F_{+,0}(t) = 1 + \lambda_{+,0}' \frac{t}{m_\pi^2}+\frac{1}{2} \lambda_{+,0}'' \left( \frac{t}{m_\pi^2}\right)^2 +\cdots\label{Taylor}
\eeq
From fits to the $K_{l3}$ spectra one can obtain the constants $\lambda_{+,0}'$
and $\lambda_{+,0}''$. 
 The study
of $F_{+,0}(q^2)$ in \tauKpi, where $(m_K+m_\pi)^2<s\equiv q^2<m_\tau^2$ , 
 is  welcome as it can further our
knowledge of the energy dependence of the form factors.  This
can lead to a better
determination of $\lambda_{+,0}'$
and $\lambda_{+,0}''$ as well as the phase space integrals that appear in the description of  $K_{l3}$ decays  and, consequently,  to an improvement
in the determination of $|V_{us}|$.

In Section \ref{old}, we briefly review some of the results of
Ref.~\cite{we} where dispersive representations of the vector form
factor were used to fit the \tauKpi~spectrum from Belle \cite{Belle}.
We emphasise the comparison of our results with others found in the
literature. In Section~\ref{combined}, we present an exploratory study
based on a combined analysis of \tauKpi~and $K_{e3}$ spectra aimed at
better determining the phase space integrals required in $K_{e3}$
decays.

\section{The {$K\pi$} vector form factor in \tauKpi~decays}
\label{old}

The differential decay distribution for the process $\tau \to K(k)\pi(p)\nu_\tau$ can be
written as \cite{Finke}
\beq
 \frac{\rd\Gamma_{K\pi}}{\rd\sqrt{s}} \,=\, \frac{G_F^2|V_{us}|^2 m_\tau^3}
{32\pi^3s}\,S_{\mbox{\tiny EW}}\left(1-\frac{s}{m_\tau^2}\right)^{\!2}     \times\left[
\left( 1+2\,\frac{s}{m_\tau^2}\right) q_{K\pi}^3\,|F_+(s)|^2 +
\frac{3\Delta_{K\pi}^2}{4s}\,q_{K\pi}|F_0(s)|^2 \right]  
\label{dGamma},
\eeq
where isospin invariance is assumed   and we have summed over the two possible
decay channels $\tau^-\to\nu_\tau\overline K^0\pi^-$ and
$\tau^-\to\nu_\tau K^-\pi^0$, that contribute in the
ratio $2\hspace{-0.4mm}:\!1$ respectively. Furthermore, $S_{\rm EW}=1.0201$
\cite{Sew} is an electro-weak correction factor,
$\Delta_{K\pi}\equiv m_K^2-m_\pi^2$, $s=(k+p)^2$, and $q_{K\pi}$ is the kaon momentum in
the rest frame of the hadronic system,
\beq
q_{K\pi}(s) \,=\, \frac{1}{2\sqrt{s}}\sqrt{\Big(s-(m_K+
m_\pi)^2\Big) \Big(s-(m_K-m_\pi)^2\Big)} \times \theta\Big(s-(m_K+m_\pi)^2\Big) .
\eeq
 In Eq.~(\ref{dGamma})  the prevailing contribution is given by
 $F_+(s)$. Note that since the $K\pi$ pair is in the final state, we
 now deal with the crossing-symmetric version of Eq.~(\ref{FF}) which
 corresponds to an analytic continuation of $F_{+,0}(q^2)$ to the region 
 $q^2\geq s_{K\pi}=(m_K+m_\pi)^2$, where the form factors develop
 imaginary parts. This renders the approximate description given by
 Eq.~(\ref{Taylor}) useless and, hence, one has to resort to more
 sophisticated treatments.  The Belle collaboration~\cite{Belle}
 employed form-factors based on Breit-Wigner expressions to describe
 the effect of resonances, among which the $K^*(892)$ largely
 dominates.  In Ref.~\cite{JPP}, a description of $F_+(s)$ based on
 resonance chiral theory (RChT) was employed and, from fits to the
 Belle spectrum, the Taylor expansion as well as the masses and widths
 of the lowest vector resonances were determined. Finally, in
 Ref.~\cite{we} we have introduced several  dispersive representations for
 $\tilde F_+(q^2)$.
 
The purpose of our study was twofold. First, from general principles of
analyticity the form factors must fulfil a dispersion relation.
Although in Ref.~\cite{JPP} the deviations from the analytic behaviour are
only small corrections of order $p^6$ in the chiral expansion, it is
interesting to corroborate this pattern by the use of a dispersive representation for $F_+(s)$. Second, a three-times-subtracted dispersive
representation of the type used in Ref.~\cite{PG}  enables us to produce less model dependent results. To make the
argument clearer let us quote the expression of $\tilde F_+(s)$ used
in our best fit~\cite{we}
\beq
\label{Ftil3sub1res}
\tilde F_+(s) \,=\, \exp\left [ \alpha_1\, \frac{s}{m_{\pi^-}^2} +
\frac{1}{2}\alpha_2\frac{s^2}{m_{\pi^-}^4}   + \frac{s^3}{\!\pi}
\int\limits^{s_{\rm cut}}_{s_{K\pi}} \!\!ds'\, \frac{\delta_1^{K\pi}(s')}
{(s')^3(s'-s-i0) }\right] \,.\label{dispFF}
\eeq
In the last equation, the two subtraction constants $\alpha_1$ and
$\alpha_2$ are obtained from a fit to the Belle spectrum. These
constants are related to the Taylor expansion~(\ref{Taylor}) as
$\lambda_+'=\alpha_1$ and
$\lambda_+''=\alpha_2+\alpha_1^2$. Concerning the phase
$\delta_1^{K\pi}(s)$, up to the first inelastic threshold unitarity
ensures that $\delta_1^{K\pi}(s)$ is the $K\pi$ $P$-wave scattering
phase shift. For simplicity, in Eq.~(\ref{dispFF}) we consider only the
$K\pi$ channel. An advantage of the three-times-subtracted form of
$F_+(s)$ is the fact that the integral over the phase is highly
suppressed by the factor $(s')^3$ in the denominator of the
integrand. Therefore, the high-energy portion of $\delta^{K\pi}_1$
weights little, laying emphasis to the elastic domain for which we can
provide a reliable model.  We vary the cut-off $s_{cut}$ in the interval
$(1.8 \,\, \mbox{GeV})^2<s_{cut}< \infty$ to quantify this
suppression.

In practice, when using Eq.~(\ref{dispFF}) one needs a functional form
for the phase $\delta^{K\pi}_1(s)$. We take a form inspired
by the RChT description of Ref.~\cite{JPP}. The phase  reads
\beq\label{phase}
\delta_1^{K\pi}(s) = \tan^{-1}\left[\frac{\IM\, \tilde F_+(s)}{\RE\, \tilde  F_+(s)}\right],
\eeq
where
\begin{equation}
\label{FpKpi2}
\tilde F_+(s) \,=\, \frac{m_{K^*}^2 - \kappa_{K^*}\,\tilde H_{K\pi}(0) +
\gamma \, s}{D(m_{K^*},\gamma_{K^*})} -
\frac{\gamma\, s}{D(m_{K^{*'}},\gamma_{K^{*'}})} \,.
\end{equation}
In the last equation,   the first piece on  the right-hand side corresponds to the $K^*(892)$ whereas
the second accounts for the $K^*(1410)$.  The parameter $\gamma$ is obtained from fits to data and $\tilde H_{K\pi}(s)$ is the one-loop integral (for its precise definition we refer to Ref.~\cite{JPP}).
The denominators are given by
\begin{equation}
\label{Den}
D(m_n,\gamma_n) \,\equiv\, m_n^2 - s - \kappa_n \, \RE\,\tilde H_{K\pi}(s) -
i\, m_n \gamma_n(s) \,,
\end{equation}
where the constants $\kappa_n$ are defined so that $-i \kappa_n\, \IM\, \tilde H_{K\pi}(s) = - i m_n \gamma_n(s)$ and the running width of a vector resonance is taken to be 
\beq
\gamma_n(s) = \, \gamma_{n}\frac{s}{m_{n}^2} \frac{\sigma^3_{K\pi}(s)}
{\sigma^3_{K\pi}(m_{n}^2)} \,,
\eeq
with $\gamma_n\equiv\gamma_n(m_n^2)$ and $\sigma_{K\pi}(s)=2 q_{K\pi}/\sqrt{s}$. The parameters $m_n$ and $\gamma_n$ are determined by the fit.  Concerning the phase
(\ref{phase}), the main difference as compared with that of Ref.~\cite{JPP}
is that the real part of $\tilde H_{K\pi}(s)$ is resummed into the
functions $D(m_n,\gamma_n)$. This procedure shifts the values of our
parameters $m_n$ and $\gamma_n$ with respect to the ones of
Refs.~\cite{Belle, JPP}. As we show  below,  the {\it physical} pole position of the resonances are not affected by this shift.

Although the main contribution to the decay \tauKpi~is given by the
vector form factor, the scalar component in Eq.~(\ref{dGamma}) cannot
be neglected.  A comprehensive coupled-channel description of $F_0(s)$
in the RChT framework plus dispersive constraints was given in
Ref.~\cite{F0}. Here, for $F_0(s)$ we take the last numerical update of
Ref.~\cite{F0Last}. Finally, in order to compare Eq.~(\ref{dGamma}) with real data,
one needs an ansatz for the number of events in the $i$-th bin  with centre
at $b^i_c$ and width $b_w$.  The theoretical number of events is then
\beq
\label{Nth}
N(i) =\mathcal{N}_T\,  \frac{1}{2}\frac{2}{3}  b_w\,  \frac{1}{\Gamma_\tau \,\, \bar B_{K\pi}}  \frac{\rd\Gamma_{K\pi}}{\rd\sqrt{s}} (b_c^ i),
\eeq
where the factors $1/2$ and $2/3$ are introduced to take into account
that the $K_S\pi^-$ channel was analysed, $\mathcal{N}_T$ is the total
number of events, $\Gamma_\tau$ is the total $\tau$-lepton decay width
and $\bar B_{K\pi}$ is a normalisation that for a perfect agreement
between the data and the model would be the branching fraction
$B_{K\pi}=\mathcal{B}(\tau\to K_S\pi^-\nu_\tau)$.

\begin{figure}[!ht]
\begin{center}
\includegraphics[width=0.49\columnwidth,angle=0]{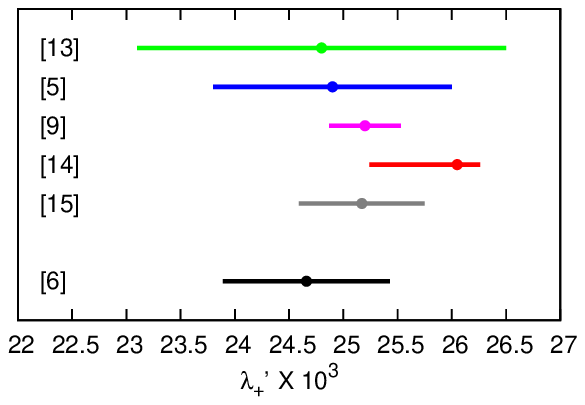}
\includegraphics[width=0.49\columnwidth,angle=0]{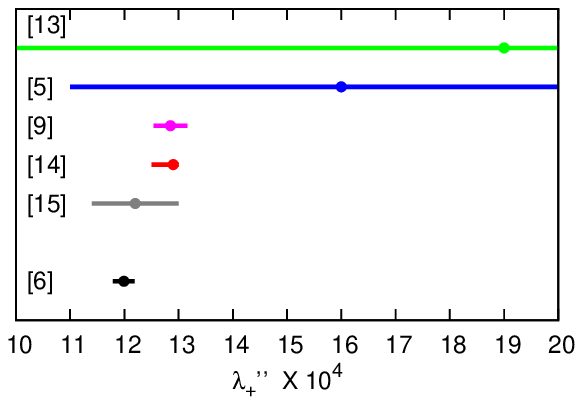}
\caption{{Values for $\lambda_+'$ (left) and $\lambda_+''$ (right) from Refs.~\cite{Flavianet, JPP, PDG,  Mou, Emilie}} compared to the ones from Ref.~\cite{we}.
  } 
\label{Fig1}
\end{center}
\end{figure}

The best fit of Ref.~\cite{we} is obtained using Eq.~(\ref{dispFF})
for $\tilde F_+(s)$, Eq.~(\ref{phase}) for $\delta_1^{K\pi}(s)$ and
the ansatz~(\ref{Nth}).  First, let us  compare our results~\cite{we}
\beq
\lambda_+' =  (24.66 \pm 0.77) \times 10^{-3}   , \qquad \lambda_+''=(11.99 \pm 0.20) \times 10^{-4} \,,
\eeq
 with 
other recent determinations of these two constants  found
in Refs.~\cite{Flavianet, JPP, PDG, Mou, Emilie}. These values are compared in  Fig.~\ref{Fig1}. 
 From Refs.~\cite{Flavianet, PDG} we quote the
results from the quadratic fit to $K_{l3}$ data. The results from Ref.~\cite{JPP} are
obtained from a fit to the Belle data set for $\tau\to
K_S\pi^-\nu_\tau$, as already commented.
In Ref.~\cite{Mou} a
coupled-channel dispersive representation constrained by scattering
data was employed whereas in Ref.~\cite{Emilie} a different
single-channel dispersive representation was used to analyse $K_{e3}$
data from the KTeV collaboration.

 It emerges from Fig.~\ref{Fig1} that the determinations of
$\lambda_+'$ are in agreement. However, the results obtained from
quadratic fits~\cite{Flavianet, PDG}, shown as the first and second
entries, display  larger uncertainties. The use of dispersive
representations, as in Refs.~\cite{we, Mou, Emilie}, the data for
\tauKpi~\cite{JPP}, or both~\cite{we}, significantly reduces the
uncertainty.  The pattern repeats itself for $\lambda_+''$, but now
the uncertainties in the case of Refs.~\cite{we, JPP, Mou, Emilie} are
impressively smaller. This comparison reveals the potential of using
dispersive representations for $F_+(s)$ and especially if combined with
	the \tauKpi~data.

\begin{figure}[!ht]
\begin{center}
\includegraphics[width=0.49\columnwidth,angle=0]{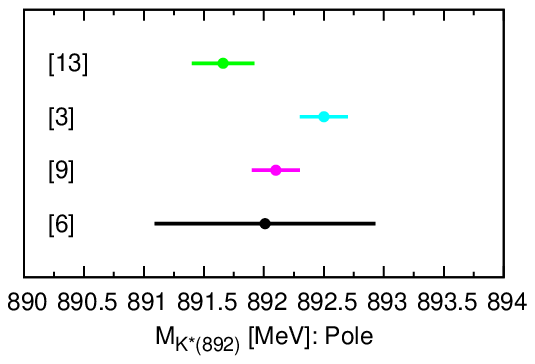}
\includegraphics[width=0.49\columnwidth,angle=0]{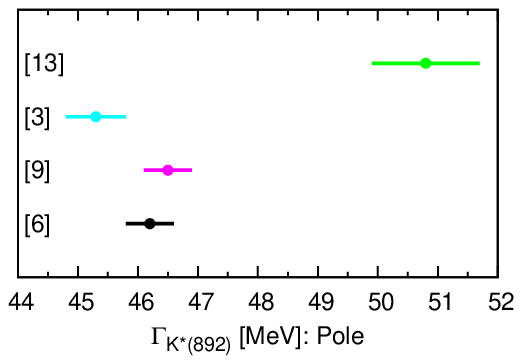}
\caption{{Mass (left) and width (right) of the charged $K^*(892)$ as defined  from the pole position (see text) calculated for Refs.~\cite{Belle, JPP, we} compared with the value from the PDG~\cite{PDG}.  For
Ref.~\cite{Belle} we employed the values of the second fit of their Table 4.1 and quote solely statistical uncertainties.
  } }
\label{Fig2}
\end{center}
\end{figure}

Since in our description the phase of $F_+(s)$ is determined from the
data, we are able to produce new values for the resonance pole
positions. In this context, it is fundamental to distinguish between
the physical pole position in the second Riemann sheet and the
parameters $m_n$ and $\gamma_n$ in Eq.~(\ref{Den}). In fact, the
parameters depend strongly on the specific form of Eq.~(\ref{Den}).
On the contrary, the poles that arise from different models are
compatible since they represent the most model independent definition
of a resonance~\cite{Rafel}. For the $K^*(892)$, we have shown (see Table 5.1 of Ref.~\cite{we}) that
although $m_{K^*}$ and $\gamma_{K^*}$ from Refs.~\cite{Belle, JPP, we} are
rather different, the physical mass and width defined from the pole
position $s_p$ as $\sqrt{s_p}=M_R -(i/2)\Gamma_R$ are compatible. They
are compared to the value quoted by the PDG~\cite{PDG} in
Fig.~\ref{Fig2}. The values obtained from \tauKpi~point towards a
smaller    width.

\section{Fitting  \tauKpi~and  $K_{e3}$ spectra}
\label{combined}

As already mentioned, in the benchmark extraction of $|V_{us}|$ from
$K_{l3}$ decays the precise knowledge of the energy dependence of 
$F_{+,0}(s)$ is important to obtain reliable values for the phase
space integrals. This problem is simplified in the case of $K_{e3}$
decays, where the lepton in the final state is an electron.
The scalar form
factor contribution is suppressed by the square of the
electron mass and, hence, the result is dominated by
$F_+(s)$. 
We denote
the phase space integral for the process $K^0\to \pi^- e^+ \nu_e $ as
$\IK$ whose expression can be found in Ref.~\cite{LR}.

Using  the best fit of Ref.~\cite{we}, we obtain the
following value for the integral: $\IK=0.15420(42)$. This result can be compared
with the one quoted by the Flavianet Kaon WG in Ref.~\cite{Flavianet} from
the average of quadratic fits: $\IK=0.15457(29)$. Although they are
compatible, the latter is more precise in spite of the fact that the
input $\lambda_+'$ and $\lambda_+''$ have larger uncertainties (see
Fig.~\ref{Fig1}). This is due to the correlation
$\rho(\lambda_+',\lambda_+'' )$ between the parameters in the two
fits. In the quadratic fit to $K_{e3}$ decays, $\lambda_+'$ and
$\lambda_+''$ turn out strongly anti-correlated
$\rho(\lambda_+',\lambda_+'' )_{K_{e3}} =-0.95$ whereas in our fit
to \tauKpi~the correlation is large and positive $\rho(\lambda_+',\lambda_+''
)_{\tau} =0.926$.

We have performed an exploratory study in order to determine whether a
combined analysis of \tauKpi~and $K_{e3}$ data could yield a more
precise result for $\IK$. For want of a true data set for the $K_{e3}$
decays we made use of a simulation aimed at reproducing the situation
of KLOE's data analysis~\cite{KLOE}. To that end, using the
expressions of Ref.~\cite{LR}, we constructed an ansatz for the number of
events similar to that of Eq.~(\ref{Nth}).  Then, assuming that the
number of events follow a Poisson distribution, we generated
$7.5\times 10^{5}$ events that were split into 30 histograms. 
A  quadratic fit to the generated data set yielded  results very similar to
the ones of Ref.~\cite{KLOE}.

\begin{table}[!ht]
\begin{center}
\caption{Main results of the simultaneous fit using Eq.~(\protect\ref{dispFF}) for $F_+(s)$. See text for details.}
\vspace{2mm}
\begin{tabular}{c  c}
\hline
\hline
 $\lambda_+'\times 10^{3}$ & $25.10 \pm (0.43)_{fit}\pm (0.07)_{s_{cut}} $  \\
 $\lambda_+''\times 10^{4}$ & $12.13 \pm (0.17)_{fit}\pm (0.13)_{s_{cut}} $   \\
$\bar B_{K\pi} [\%]$  &  $0.430\pm (0.014)_{fit}\pm (0.005)_{s_{cut}}  $         \\
\hline
$(B_{K\pi}) [\%]$       &    0.427 \\
 $\rho( \lambda_+',\lambda_+''  )$  & 0.845  \\
$\chi^2/d.o.f.$  & 427/438  \\
\hline
\hline
\end{tabular}
\label{Tab1}
\end{center}
\end{table}

With this data set, we carried out a simultaneous fit of the generated 
$K_{e3}$ data and the Belle spectrum of \tauKpi~using our
representation, Eq.~(\ref{dispFF}), for $F_+(s)$. The main results are shown in
Table~\ref{Tab1} where we explicitly indicate the systematic error due to
$s_{cut}$.  Since the parameters are much better constrained at
low-energies, it is possible to keep the normalisation $\bar B_{K\pi}$
in Eq.~(\ref{Nth}) as a free parameter. The result obtained from the
fit is compatible with the world average  $\mathcal{B} = 0.418 \pm
0.011\%$~\cite{BKpi} and the integrated value, denoted $(B_{K\pi})$ in
Tab.~\ref{Tab1}, is very close to $\bar B_{K\pi}$. Furthermore, in
this fit the uncertainty in $\lambda_+'$ is reduced and, more
important, is mainly driven by statistics.
From the results of this fit we obtain $\IK= 0.15444(24)$, which has
a smaller uncertainty than the result of Ref.~\cite{Flavianet}. Of course,
we are by no means recommending this value, since it is based on a
simulated data set. However, it is clear that the prospects are very
positive since the statistics for \tauKpi~will soon be improved with the
forthcoming spectrum from the BaBar collaboration~\cite{BKpi} thus
reducing the uncertainty even further.


\vspace{-3mm}


\acknowledgments
\vspace{-2mm}
  This work has been supported in
  part by the  {\em Ramon y Cajal } programme (R. Escribano), {\em Ministerio de  Ciencia e Innovaci\'on} under grant
  CICYT-FEDER-FPA2008-01430, the EU Contract No. MRTN-CT-2006-035482
 ``FLAVIAnet", the Spanish Consolider-Ingenio 2010 Programme CPAN
  (CSD2007-00042),  and the {\it
   Generalitat de Catalunya} under grant 2005-SGR-00994.
\vspace{-1mm}




\begin{thebibliography}{99}
 
 \bibitem{alphas}{E.~Braaten}, {S.~Narison}, and {A.~Pich},
\newblock { Nucl. Phys. B} {\bf 373} (1992) 581; { M.~Davier} {\it et al.}, 
\newblock { Eur. Phys. J. C} {\bf 56} (2008) 305
\newblock [arXiv:0803.0979  [hep-ph]]; { M.~Beneke} and { M.~Jamin},
\newblock { JHEP} {\bf 09} (2008) 044 [arXiv:0806.3156  [hep-ph]]; K.~Maltman and T.~Yavin,
  Phys.\ Rev.\  D {\bf 78} (2008) 094020
  [arXiv:0807.0650  [hep-ph]].

 
 \bibitem{msVus}{E.~G\'amiz} {\it et al.},
\newblock { Phys. Rev. Lett.} {\bf 94} (2005) 011803 [arXiv:hep-ph/0408044];  
{ JHEP} {\bf 01} (2003) 060 [arXiv:hep-ph/0212230].





 \bibitem{Belle} D.~Epifanov {\it et al.}  [Belle Collaboration],
  Phys.\ Lett.\  B {\bf 654} (2007) 65
  [arXiv:0706.2231 [hep-ex]].


 
 \bibitem{LR} H.~Leutwyler and M.~Roos,
  Z.\ Phys.\  C {\bf 25} (1984) 91.

 
 \bibitem{Flavianet} M.~Antonelli {\it et al.}  [FLAVIAnet Working Group on Kaon Decays],
  [arXiv:0801.1817 [hep-ph]].

 
 \bibitem{we} D.~R.~Boito, R.~Escribano, and M.~Jamin,
  Eur.\ Phys.\ J.\  C {\bf 59} (2009) 821
  [arXiv:0807.4883 [hep-ph]].

\bibitem{Finke}   M.~Finkemeier and E.~Mirkes,
  Z.\ Phys.\  C {\bf 72}  (1996) 619
  [arXiv:hep-ph/9601275].
  
  \bibitem{Sew}{ J.~Erler},
\newblock {Rev. Mex. Fis.} {\bf 50} (2004) 200,
\newblock [arXiv:hep-ph/0211345].



\bibitem{JPP}{ M.~Jamin}, { A.~Pich}, and { J.~Portol\'es},
\newblock { Phys. Lett.} {\bf B~640} (2006) 176
\newblock [arXiv:hep-ph/0605096]; 
  Phys.\ Lett.\  B {\bf 664} (2008) 78
  [arXiv:0803.1786  [hep-ph]].

 

\bibitem{PG}  A.~Pich and J.~Portoles,
  Phys.\ Rev.\  D {\bf 63} (2001) 093005
  [arXiv:hep-ph/0101194].

%



\bibitem{F0}  M.~Jamin, J.~A.~Oller, and A.~Pich,
  Nucl.\ Phys.\  B {\bf 622} (2002) 279  [arXiv:hep-ph/0110193]; \\ 
 Eur.\ Phys.\ J.\  C {\bf 24} (2002) 237  [arXiv:hep-ph/0110194]; 
  JHEP {\bf 0402} (2004) 047 [arXiv:hep-ph/0401080].
 

\bibitem{F0Last} M.~Jamin, J.~A.~Oller, and A.~Pich,
  Phys.\ Rev.\  D {\bf 74} (2006) 074009
  [arXiv:hep-ph/0605095].


\bibitem{PDG} C.~Amsler {\it et al.}  [Particle Data Group],
  Phys.\ Lett.\  B {\bf 667}  (2008) 1.


\bibitem{Mou} B.~Moussallam,
  Eur.\ Phys.\ J.\  C {\bf 53} (2008) 401
  [arXiv:0710.0548 [hep-ph]].


\bibitem{Emilie}V.~Bernard {\it et al.},
  [arXiv:0903.1654 [hep-ph]]; E.~Abouzaid {\it et al.}, in
preparation.




\bibitem{Rafel} { R.~Escribano} {\it et al.}, 
\newblock { Eur. Phys. J. C} {\bf 28} (2003) 107
\newblock [arXiv:hep-ph/0204338].


\bibitem{KLOE} F. Ambrosino {\it et al.} [KLOE Collaboration], Phys. Lett. B {\bf 636} (2006) 166 [arXiv:hep-ex/0601038].

\bibitem{BKpi}A. Wren, talk presented at Tau08, Novosibirsk, Russia; { B.~Aubert} { et~al.} [BaBar Collaboration],
\newblock talk presented at ICHEP08, Philadelphia, Pensylvania,
  [arXiv:0808.1121v2 [hep-ex]].


 

 
 
\end{thebibliography}
\end{document}